# Improved graphene blisters by ultrahigh pressure sealing

Yolanda Manzanares-Negro[1], Pablo Ares[1†*], Miriam Jaafar[1], Guillermo López-Polín[1‡*], Cristina Gómez-Navarro[1], and Julio Gómez-Herrero[1]

[1]Departamento de Física de la Materia Condensada and Condensed Matter Physics Center IFIMAC. Universidad Autónoma de Madrid, 28049, Madrid, Spain.

Present address:

†Department of Physics & Astronomy and National Graphene Institute, University of Manchester. Manchester M13 9PL, UK.

‡Instituto de Ciencia de Materiales de Madrid (ICMM), CSIC. 28049, Madrid, Spain.

*E-mail: pablo.ares@manchester.ac.uk, E-mail: guillermo.lp@csic.es

Graphene is a very attractive material for nanomechanical devices and membrane applications. Graphene blisters based on silicon oxide micro-cavities are a simple but relevant example of nanoactuators. A drawback of this experimental set up is that gas leakage through the graphene-$SiO_2$ interface contributes significantly to the total leak rate. Here we study the diffusion of air from pressurized graphene drumheads on $SiO_2$ micro-cavities and propose a straightforward method to improve the already strong adhesion between graphene and the underlying $SiO_2$ substrate, resulting in reduced leak rates. This is carried out by applying controlled and localized ultrahigh pressure (> 10 GPa) with an Atomic Force Microscopy diamond tip. With this procedure, we are able to significantly approach the graphene layer to the SiO$_2$ surface around the drumheads, thus enhancing the interaction between them allowing us to better seal the graphene-SiO$_2$ interface, which is reflected in up to ~ 4 times lower leakage rates. Our work opens an easy way to improve the performance of graphene as a gas membrane on a technological relevant substrate such as SiO$_2$.

INTRODUCTION

Even though it is only one atom thick, due to its defect-free nature, graphene shows great impermeability[1-4] and seems to be the ideal candidate to enclose even the smallest gas molecules,[3,4]. The standard setup used to exploit this extreme impermeability is based on the creation of an open micro-cavity within other bulk material (usually SiO$_2$) that is subsequently covered by a graphene layer acting as a drumhead (see Figure 1a for a schematic view). This configuration has been already proved useful for applications such as graphene microphones,[5,6] pressure sensors,[7] modifying the thermal expansion coefficient of graphene[8] or tunable systems to confine charge carriers.[9] In addition, these graphene blisters have been used in fundamental research experiments with the aim of determining mechanical magnitudes such as elastic modulus[2] and the effect of strain on them[10] or its permeability to different gases.[3]

The high impermeability of such devices relies on a very high adhesive interaction between the graphene layer and the underlying material. In a pioneer work by Bunch *et al.*,[3] the adhesion energy of monolayer graphene and SiO$_2$ was reported to be 0.45 J/m$^2$. This anomalously high figure is attributed to the extreme flexibility of graphene, which allows it to conform to the topography of the underlying material, maximizing surface adhesion. However, this upper bound value of adhesion energy is usually not reached in most samples, and lower adhesion energies and higher leakage rates are usually measured.[11,12]

Although the presence of wrinkles and folds in graphene or residue in the graphene-substrate interface are known to highly modify adhesion energy, it was not clear whether the origin of pressure equilibration in graphene blisters came from diffusion of gas molecules through the $SiO_2$ or through the graphene-$SiO_2$ interface.[2] This issue is crucial for stable measurements since the reported leakage rates would mean that a typical cavity with dimensions of 1 x 1 x 0.5 microns under a pressure difference of 4 bar would lose 50% pressure in just a few hours. Lee *et al.* have used electron beam induced deposition to place $SiO_2$ across the edge of suspended multilayer graphene flakes and saw a remarkable reduction in the permeation rate,[13] indicating that the diffusion of molecules through the graphene-$SiO_2$ interface is the predominant mechanism for gas leakage in graphene-$SiO_2$ cavities. More recently, Sun *et al.* have fabricated graphene drumheads by using electron-beam lithography and dry etching to produce micrometer-sized containers on monocrystals of graphite or hexagonal boron nitride, sealing them with graphene monolayers.[4] With this approach, they obtained atomically tight sealing because of the clean and atomically sharp interface, demonstrating the impermeability of graphene to even the smallest molecules, with maybe the exception of hydrogen due to the catalytic activity of ripples, wrinkles and other defects in graphene. Although these approaches allow for a remarkable decrease of the gas diffusion along the resulting interfaces, they rely on complex fabrication techniques and limited substrate configurations.

In this work, we study the role of the diffusion of molecules through the graphene-$SiO_2$ interface and propose a simple technique that improves significantly the cavity sealing in a widely used technological support such as $SiO_2$ by increasing the adhesion between graphene and the underlying substrate. It is based on the application of very local ultrahigh pressure (> 10 GPa) on a nanometer wide rim surrounding the cavity by means of an Atomic Force Microscopy (AFM) diamond tip.[14] With this technique, we have been able to improve the leak characteristic time of our micro-cavities up to a factor of ~ 4, showing that improving the adhesion between graphene and the underlying substrate is crucial to reduce the diffusion of molecules through the interface in these micro-cavities.

RESULTS AND DISCUSSION

We prepared graphene drumheads by mechanical exfoliation of natural graphite on $SiO_2$ (300 nm)/Si substrates with predefined circular wells of diameters ranging from 0.5 to 3 μm. We selected only flat (unfolded, unwrinkled) monolayer graphene flakes for this study. AFM images of the drumheads revealed that, as reported previously, graphene layers adhere to the vertical walls of the wells for ~ 2-10 nm in depth in their initial morphology.[15, 16] We carried out all the experiments described in this work in a variable pressure chamber equipped with an Atomic Force Microscope.[10] By controlling the internal pressure in the micro-cavity ($P_{in}$) and external pressure within this chamber ($P_{out}$), we are able to apply a pressure difference of up to 4 atm to our graphene blisters for convex geometry ($P_{out} > P_{in}$) and almost 1 atm for concave geometry ($P_{out} < P_{in}$). See Figure 1a for a schematic view. Once a pressure difference is applied through the drumhead, we acquired consecutive AFM topography images in dynamic mode in the same area for long (hours-days) periods of time (see Methods). All the experiments presented herein used $N_2$ or air as filling gas obtaining similar results.

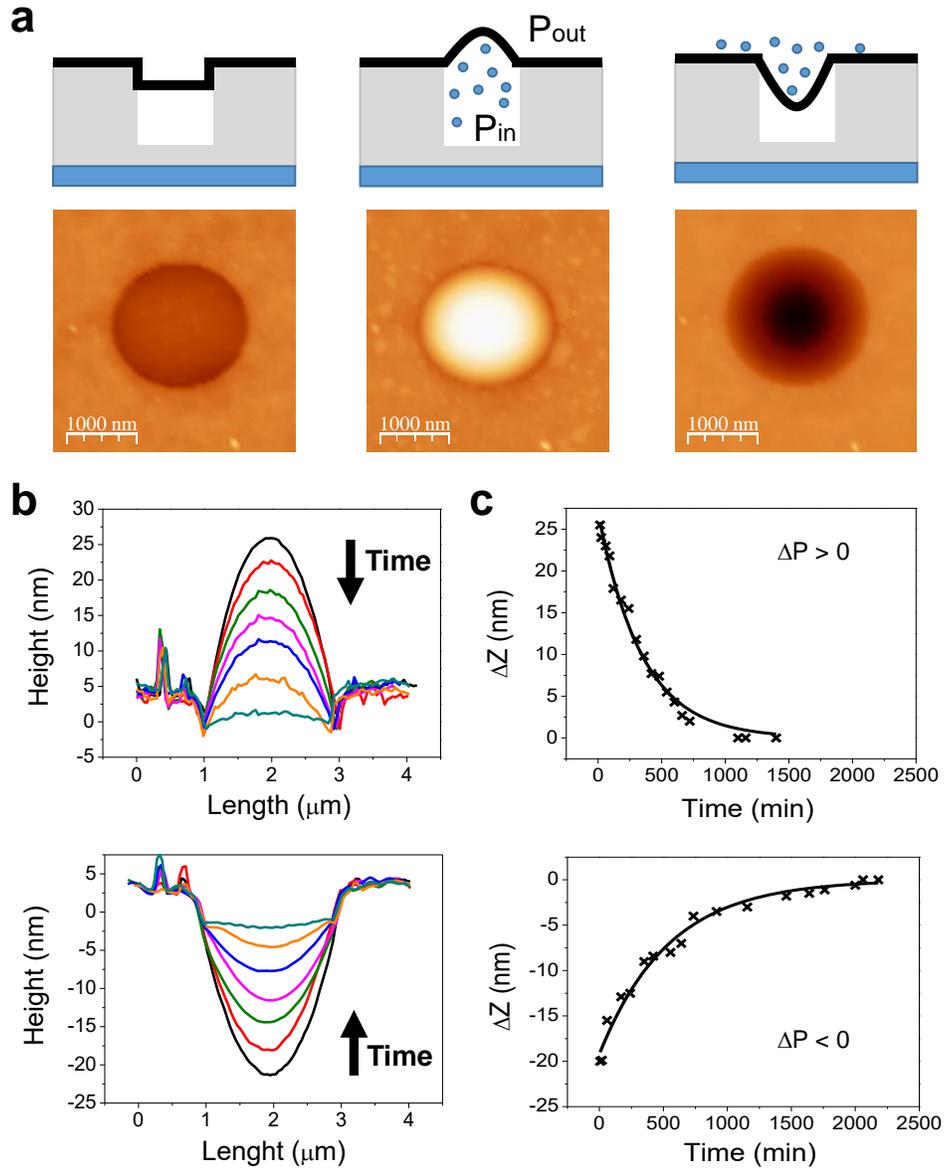

**Figure 1.** Leakage rate for two different blister configurations. (a) Schematic view (top) and corresponding AFM topographical images (bottom) of the possible configurations. (b) AFM line traces taken through the center of the graphene membrane of (a). (c) Time evolution of the blister extrema positions when the internal pressure is higher than the external, ∆P = +1 atm (top) and when the internal pressure is lower than the external, ∆P = -1 atm (bottom). Fitting to an exponential decay yields characteristic times of 345 and 540 min respectively.

Figure 1a shows a scheme of the possible configurations of the blisters according to the different combinations of $P_{in}$ and $P_{out}$ (top panel) together with representative AFM topographies (bottom panel). For a blister where the pressures are equilibrated ($P_{in} = P_{out}$) (Figure 1a, left), the flake appears flat, slightly adhered to the vertical wall of the well. In the case of a blister that initially supports an internal pressure higher than the external one ($P_{in} > P_{out}$) (Figure 1a, middle), the flake deflects outwards *i.e.* concave geometry. If the pressure is applied in the opposite direction $P_{in} < P_{out}$ (Figure 1a, right) the graphene membrane deflects inwards *i.e.* convex geometry. In what follows, we define $\Delta P = P_{in} - P_{out}$, thus $\Delta P > 0$ implies concave blister shape and $\Delta P < 0$ convex. By monitoring the maximum/minimum positions (extrema) of the blister through AFM images (see Figure 1b), we always observe that, in absolute value, they always decrease with time, implying that the pressure difference across the membrane tends to equilibrate. For these experiments it is important to remark that what we can measure from the AFM topographies is the blister shape and not the pressure inside the well. We initially leave the sample in the chamber at a certain pressure overnight in order to equilibrate the pressures on both sides of the drumhead. Then, we fix an initial $\Delta P$ by suddenly varying the pressure in the chamber and we approach the AFM tip to the surface. This process usually takes a few minutes, along this time the gas is already flowing from/to the micro-cavity, as it tends to reduce the absolute value of $\Delta P$. This implies that, when we finish the first AFM image, the blister $\Delta P$ is unknown, hence the observable is the height of the extrema of the blister. Thus, all the values for $\Delta P$ considered in this work are values for the initial pressure difference that we set in the macroscopic chamber. Importantly, *ΔP* can be expressed as a function of the maximum/minimum deflection of the blister $Z$ as[2, 17]

$$\Delta P = \frac{4}{\pi a^2}\left(c_1 S_0 Z + \frac{4 c_2 E t}{\pi a^2 (1-v)} Z^3\right) \quad (1)$$

Where $a$ is the radius of the pressurized circular region, $c_1$ = 3.393, $c_2 = (0.8 + 0.062 v)^{-3}$ with $v$ the Poisson's ratio ($\sim$ 0.16), $Et$ = 340 N m$^{-1}$ for monolayer graphene and $S_0$ is the pre-tension accumulated in the sheet.

Figure 1c presents the evolution of the extrema for the blister in Figures 1a and b, for two opposite pressure differences (+1 atm, top panel and -1 atm, bottom panel). Here we can appreciate how the extrema evolve with time towards cero (flat membrane). However, the evolution for the concave geometry is faster than for the convex one. To quantify this time we have heuristically fitted the experimental data to an exponential decay curve (see the Supporting Information for details)

$$\Delta Z = Z_0 e^{-\frac{t}{\tau}} \qquad (2)$$

Being $Z_0$ the initial extrema of the blisters, and $\tau$ a characteristic time, obtaining values of 345 min for the concave geometry ($\Delta P > 0$) and 540 min for the convex one ($\Delta P < 0$) for the same blister. The fact that $\tau$ is higher for $\Delta P < 0$ suggests that the shape of the membrane with respect to the SiO$_2$ substrate, and hence the graphene-SiO$_2$ interaction, has a marked influence on the leakage rate, in contradiction with what would be expected for a pressure equilibrium mechanism dominated by diffusion through SiO$_2$. The higher pressure outside the convex blister is pushing the entire graphene membrane against the substrate, and the elastic energy of the deformed membrane is accumulated on the edges, introducing extra effective adhesion. Consequently, the diffusion of molecules through the graphene-substrate interface is more difficult. In the concave case the behaviour is the opposite, the higher pressure inside the blister is pushing the graphene membrane out of the substrate, even causing delamination/peeling of the flake at the edges if the pressure is high enough,[17, 18] thus facilitating the diffusion of the gas molecules through the interface (see Figure S1 in the Supporting Information).

We have also observed that the leakage rate is different for different micro-cavities and from one flake to other. These variations can be attributed to a combination of different factors such as uncontrolled tensions, atomic-sized ripples or folds,[19] surface contamination and the distance of the microchamber to the nearest graphene flake edge. Whereas it is experimentally challenging to account for the contribution of these ripples and contamination, Figure 2 presents a comparison of leakage rates for different blisters of the same dimensions formed with the same graphene flake. Figure 2a portrays an optical microscopy image where a graphene flake covers the substrate forming several drumheads. Here we will focus on 3 different cavities (marked with square, circle and triangle) with the same diameter but different distances to the graphene edge. Figures 2b and c show the evolution of the extrema positions of the membrane as a function of time for the three blisters above mentioned. The symbols in the plot correspond to the polygons used in Figure 2a, leading to characteristic times for the concave geometry ($\Delta P > 0$) of $\tau \approx 175$, 340 and 365 min for squares, circles and triangles respectively and 310, 755 and 855 min for the convex one ($\Delta P < 0$). For all cases, the characteristic times are higher for $\Delta P < 0$. The different $\tau$ for each of the geometries reflects the distance from each blister to the nearest edge of the flake. This is clearly observed in Figure 2d, where we represent a plot of the characteristic times for different blisters in the flake as a function of the distance to the edge of the flake.

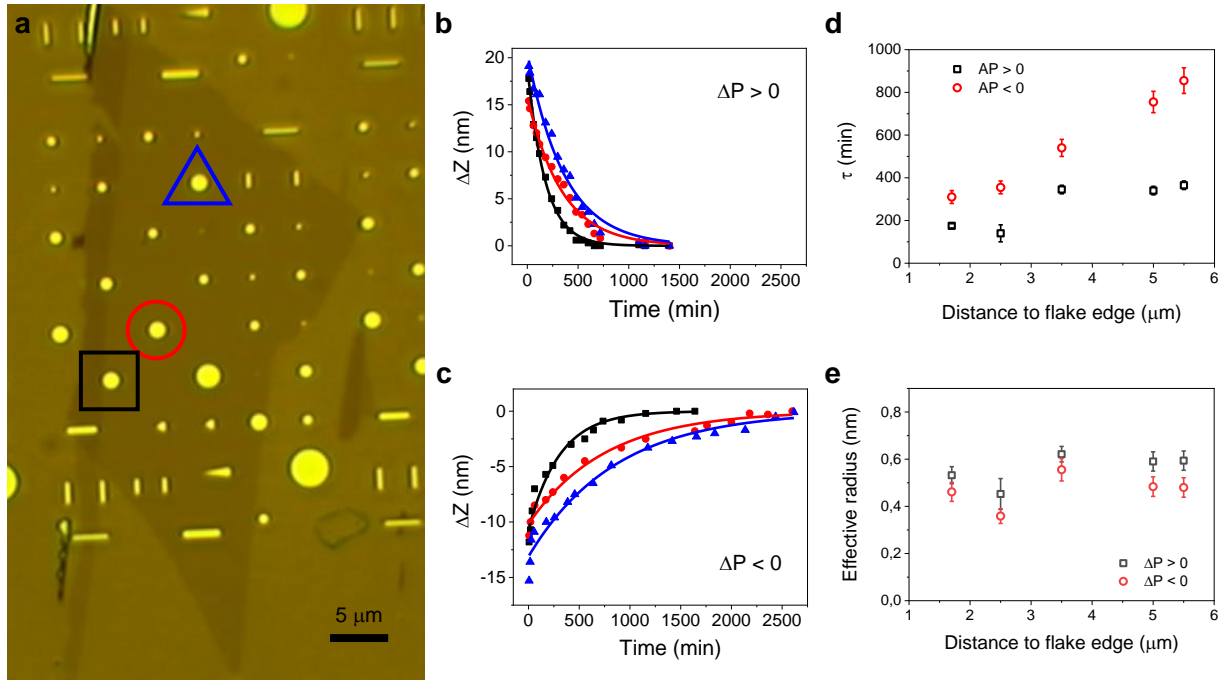

**Figure 2.** Comparison of leakage rates for different blisters. (a) Optical image of a graphene flake (mostly one layer) obtained by microexfoliation on a substrate with holes. (b) Time evolution of the blister maximum position for the holes (1.5 μm in diameter) surrounded by polygons in (a) when the initial pressure inside the microchambers is higher than the external pressure (ΔP = +1 atm). Fitting to an exponential drop yields characteristic times of 175, 340 and 365 min for square, circle and triangle data respectively. (c) Same as in (b) but now with an initial internal pressure lower than the external (ΔP = -1 atm). In this case, the minimum position of the blisters is represented and τ ≈ 310, 755 and 855 min for square, circle and triangle data are found respectively. (d) Characteristic times as a function of the distance to the nearest flake edge for several blisters in (a). The times for the convex configurations are always higher than the concave ones and, in both cases, the time increases with the distance to the flake edge. (e) Effective radii of the leakages from the same blisters as in (d). For a given flake, the effective radius depends on the adhesion of the flake to the substrate but not on the distance to the flake edge, as can be seen.

To gain further insight on the mechanisms governing the diffusion of gas from the pressurized blisters, we have conducted an analysis of the leakage rate. It is instructive to approach the problem by considering the analogy where the entire pressure drop takes place through a long cylindrical tube of constant cross section connecting the cavity to the outside (Figure S2 in the Supporting Information). This approximation allows us to use the Poiseuille's law considering a compressible gas[20, 21] to find an expression for the evolution of the deflection of the blisters with time, which leads to equation (2) (see Supporting Information section SI2 for the details of the analysis). This approach leads to a characteristic time given by the expression

$$\tau = \frac{16\eta MV}{\pi \rho RT} \frac{L_{eff}}{r_{eff}^4} \qquad (3)$$

Where $\eta$ is the dynamic viscosity of the fluid ($1.85 \times 10^{-5}$ kg m$^{-1}$ s$^{-1}$ for air at 298 K), $M$ the molar mass of the gas (~ 0.029 kg mol$^{-1}$ for air), $V$ the volume, $\rho$ the gas density (1.18 kg m$^{-3}$ for air at 298 K), $R$ the gas constant, $T$ the temperature, $L_{eff}$ the distance of the blister to the flake edge and $r_{eff}$ the effective radius of the tube that yields the observed leakage rate. This toy model yields valuable but still intuitive information of the system. At first glance, this analysis accounts for the linear increase of the blisters gas leakage times with the distance to the flake edge, as observed experimentally (Figure 2d). From the characteristic times obtained from the fittings we can estimate the effective radius of leakage. Figure 2e shows the effective leakage radii for the blisters in Figure 2d. As can be seen, despite the clear dependence of the characteristic times with the distance to the flake edge, the effective leakage radius does not change significantly with the distance. As all the blisters correspond to the same flake, we assume that the interaction of the membrane with the substrate is basically the same for all them. On the contrary, the leak rate can be rather different for different flakes. Figure S3 (see Supporting Information section SI3) presents the characteristic times and effective leakage radius obtained for blisters of similar characteristics (volume and distance to the edge) in five different flakes, showing significant differences between flakes. The origin of this disparity might be on uncontrolled variations of the graphene-$SiO_2$ adhesion between different flakes.

In order to improve the sealing of the micro-cavities, we have developed a method based on applying ultrahigh pressure with an AFM diamond tip[14] (see Methods). This method takes advantage of the extremely high breaking strength and flexibility of graphene that allows applying ultrahigh pressure without damaging neither the flake nor the substrate.[15, 22] It involves the following steps, as depicted in Figure 3: first, we carry out a non-invasive gentle AFM image of the flake in amplitude modulation dynamic mode (Figure 3a). Then, we apply a load corresponding to a pressure of about 25-30 GPa on a selected area with the shape of a nanometric wide square rim enclosing the desired drumhead (see a schematic view in Figure 3b). For 1.5 µm diameter drumheads, we typically produced rims with lateral dimensions for the external side 2.5-3.5 µm, internal side 1.5-2.5 µm, and widths of the modified region about 0.4-0.7 µm (see Figure S4 in the Supporting Information for images of different sealed drumheads). We made one side at a time, scanning the areas to modify in the slow scan direction twice under these ultrahigh pressure conditions. Finally, we go back to dynamic mode conditions and acquire a new topography image (Figure 3c). Comparing both images, we can clearly distinguish the depression on graphene created during the pressure load. For these pressures, a depth of ~1 nm is observed, as shown in Figure 3d corresponding to the profile along the green line in Figure 3c.

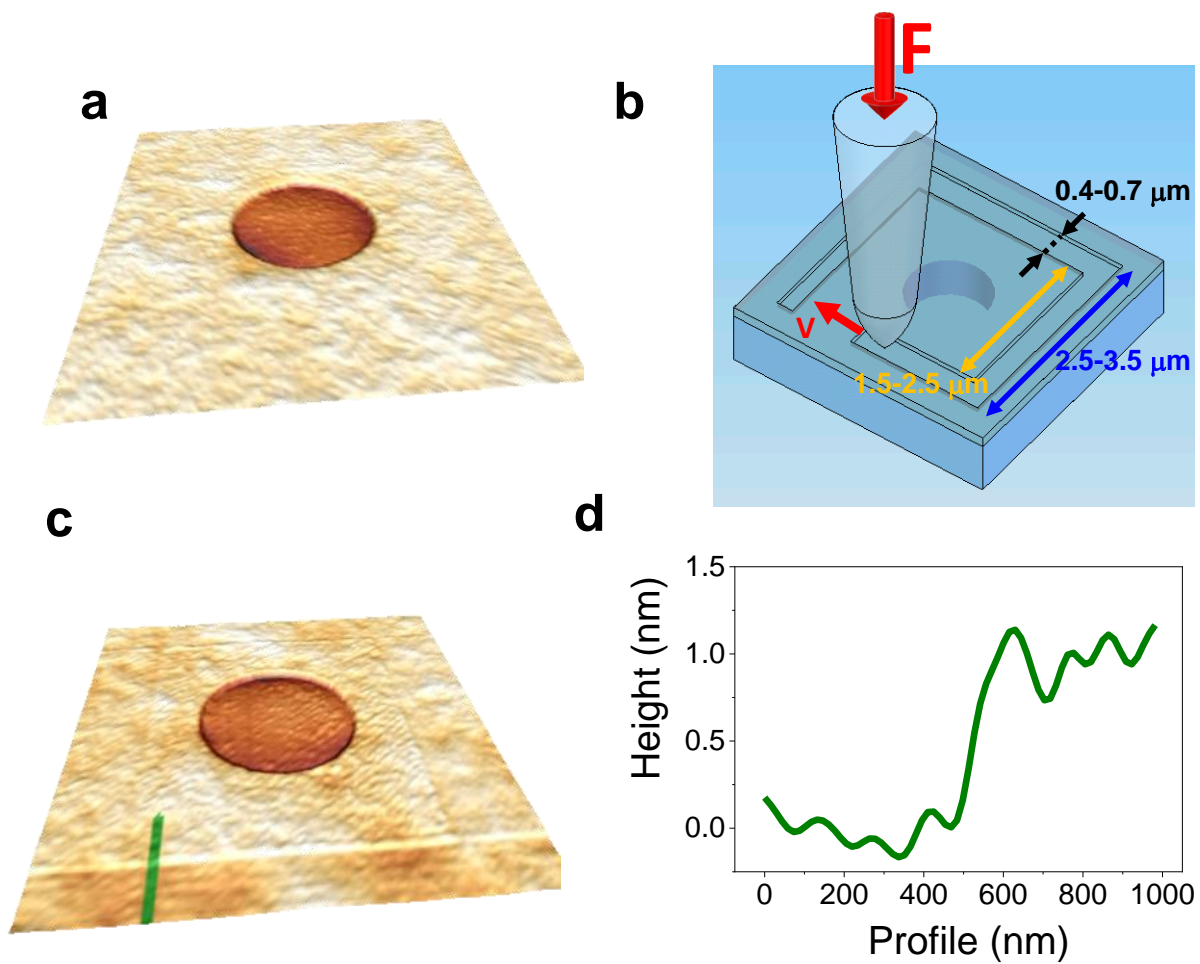

**Figure 3.** Sealing blisters. (a) AFM topography showing a 1.5 μm diameter cavity covered by a monolayer graphene flake. (b) Schematic of the process to improve the sealing: a diamond tip scans in hard contact the flake enclosing the blister in an area with the shape of a nanometric wide square rim with typical lateral dimensions as shown in the image. (c) AFM topography showing the result of the sealing process. (d) Profile along the green line in (c). The depth of the depressed region is about 1 nm.

A detailed analysis of the height differences upon pressurizing with a diamond tip was carried out in our previous work.[14] In summary, under ambient conditions there exists a layer of buffer molecules (typically water and hydrocarbons) which remains captured between the flakes and $SiO_2$. Modifications with pressures below 16 GPa did not produce any significant change on the scanned areas, suggesting that the deformation is completely reversible and that the removed buffer layer can refill the volume beneath graphene. It is important to remark that, as well as no change in height, for this pressure range there was no material accumulation on the sides of the modified areas, indicating that the effects observed with the ultrahigh pressure procedure are not related to indenting residues on top of the graphene membrane. In order to understand the effects of ultrahigh pressure on the graphene-$SiO_2$ interface, we carried out Density Functional Theory (DFT) simulations that combined with the experimental observations suggested the following scenarios.[14] For pressure values in the 16–25 GPa range, some very few covalent bonds between graphene and $SiO_2$ could have formed, which combined with the roughness of the surface, allows a limited portion of molecules to return to the interstitial region. For higher pressures ($\geq$ 25 GPa), the buffer molecules are expelled from the modified areas, a few more covalent bonds might have formed and the graphene flake remains in contact with the $SiO_2$ substrate. The depression in height in the modified area can thus be attributed to an approach of the membrane to the substrate, producing an increase of its degree of conformation and, at the same time, inducing the onset of covalent bonding between graphene and the underlying $SiO_2$ substrate.[14] Both the degree of conformation and the covalent bonding have been indeed identified as the two key aspects to increase even more the outstanding adhesion between graphene and an underlying $SiO_2$ substrate.[23] We use pressures of ~ 25-30 GPa for the sealing procedure, as such pressures produce the graphene flake to remain in contact with the $SiO_2$ substrate but are relatively far from the pressure at which the graphene sheets break.

We have applied this procedure to different blisters both for concave and convex configurations. We have chosen blisters that in the standard configuration presented characteristic times typically around 60 minutes, allowing carrying out the sealing experiments in a normal measurement session. Figure 4 presents the evolution of the extrema for representative graphene drumheads before and after sealing. While the characteristic time for a blister subjected to an initial $\Delta P = +1$ atm was $\tau = 40$ min, after sealing this time increased to 75 min (Figure 4a). For a blister with a pressure difference $\Delta P = -4$ atm (Figure 4b), the characteristic time increased from 40 to 140 min, almost by a factor of 4. Figure 4c and Figure 4d show a summary of the characteristic times and effective leakage radii for five blisters under different configurations before and after sealing. It is worth to remark that we have been able to increase the characteristic leakage time both in concave and convex configurations.

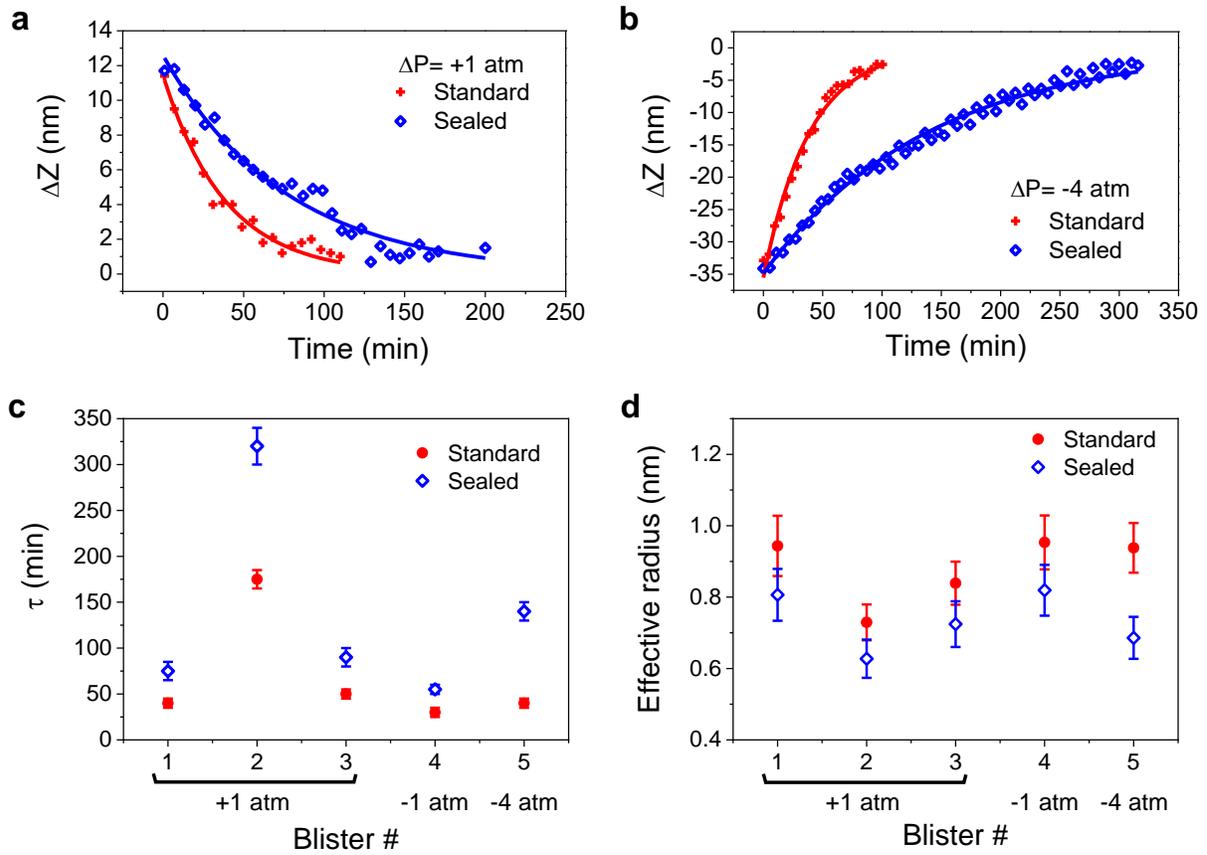

**Figure 4.** Comparison of leakage rates before and after sealing. (a) Time evolution of the maximum position for a blister with $\Delta P > 0$ in the standard (unsealed, crosses) and sealed (diamonds) cases. The characteristic times are $\tau$ = 40 and 75 min respectively. (b) Same as in (a) but now for a blister with $\Delta P < 0$. The characteristic times in this case are $\tau$ = 40 and 140 min for the standard and sealed cases respectively. (c) Summary of the change in characteristic times and (d) effective leakage radii for five blisters under different conditions. An increase of the characteristic leakage time can be seen in all the cases, which is associated to a decrease in the effective leakage radius upon ultrahigh pressure sealing.

To understand the effect of sealing under ultrahigh pressures, we can see this as an analogy to the diffusion of gas molecules through channels of different length and width. The flow rate will increase with the cross section of the channel and will decrease with its length (see Supporting Information, section SI2). Following this analogy, if the distance to the graphene edge increases, it will increase the time for the molecules to cross the channel, as observed experimentally (Figure 2d). Thus, sealing a region enclosing the blister would be equivalent to add a narrowing into the channel, being the area of the sealed region analogous to the width of the narrowing. In this case, as not all the molecules will be able to pass at the same time through the narrowing, it will increase the time for the molecules to cross the channel. Sealing a larger area will be equivalent to making the narrowing larger, thus increasing further the time. If we make the narrowing narrower, which would be equivalent to seal the area around the blister at a higher pressure, the time would be even higher. See Figure S2 in the Supporting Information for a graphic interpretation of this analogy.

Finally, we have qualitatively evaluated the increase of the graphene-substrate interaction achieved by our sealing procedure. We have indented two different drumheads, one standard and one sealed with similar characteristics in a same graphene flake, using a tip with a very high radius (250 nm). See Supporting Information section SI5 and Figure S5 for details. These especial tips allow us to achieve high average strains necessary to overcome the flake-substrate interaction without breaking the membranes, as the applied pressure at the tip-membrane contact region (P = F/A) is low. For high enough indentations, the curves present a hysteretic behavior, related to the slippage of the membrane edges during the indentation process. The hysteresis onset occurs at forces almost 3 times higher for the sealed case, evidencing a much higher membrane-substrate interaction upon ultrahigh pressure sealing.

CONCLUSIONS

To sum up, we measure the leakage rate for graphene blisters inferring that the main contribution to this magnitude is the gas flow through the graphene/substrate interface. We then demonstrate that the leakage rate can be substantially reduced by sealing the blisters using an AFM diamond tip to selectively flatten areas around the cavity rim. By doing this, we increase the graphene degree of conformation to the substrate and trigger graphene-$SiO_2$ covalent bonding, thus enhancing the adhesion between the graphene membrane and the substrate. This method opens a new simple way for the easy improvement of devices and membrane applications based on graphene on $SiO_2$ substrates.

METHODS

**Atomic Force Microscopy (AFM) imaging**. We carried out AFM measurements using a Cervantes Fullmode AFM from Nanotec Electronica SL. The AFM was located inside a variable pressure chamber (~ $10^{-6}$ mbar to 3 atm), achieving a maximum pressure difference of ~ 4 atm.[10] We employed WSxM software (www.wsxm.es) for both data acquisition and image processing.[24,25] In order to avoid sample damage, we took topographic images before and after modifications in amplitude modulation mode. For the images of blisters with $\Delta P > 0$ we used the drive amplitude modulation mode,[26] since it allows stable scanning of samples under these conditions. We employed PPP-FM cantilevers from Nanosensors (www.nanosensors.com), with a nominal resonance frequency of 75 kHz and spring constant of 2.8 N m$^{-1}$.

**Ultrahigh pressure modifications**. We employed a single crystal diamond tetrahedral pyramid SCD15/AIBS probe from MikroMasch (www.spmtips.com). The cantilever resonance frequency was 320 kHz with a spring constant of 32 N m$^{-1}$ (calibrated using the Sader method[27,28]). We calibrated the tip radius by imaging carbon nanotubes of different heights,[29] obtaining a value of R = 42 nm. We estimated the applied pressure following the Hertz's model.[30,31] More details can be found in Ref. 14.

ACKNOWLEDGMENT

We acknowledge financial support from the Spanish Ministry of Science and Innovation, through the "María de Maeztu" Programme for Units of Excellence in R&D (CEX2018-000805-M), projects PID2019-106268GB, S2018/NMT-451, FLAG-ERA JTC2017 and the Ramon Areces Foundation. G. L.-P. acknowledges financial support through the "Juan de la Cierva" Fellowship FJCI-2017-32370.

REFERENCES

1. Berry, V. Impermeability of Graphene and Its Applications. *Carbon* **2013**, *62*, 1-10.
2. Bunch, J. S.; Verbridge, S. S.; Alden, J. S.; van der Zande, A. M.; Parpia, J. M.; Craighead, H. G.; McEuen, P. L. Impermeable Atomic Membranes from Graphene Sheets. *Nano Lett.* **2008**, *8*, 2458-2462.
3. Koenig, S. P.; Wang, L. D.; Pellegrino, J.; Bunch, J. S. Selective Molecular Sieving through Porous Graphene. *Nat. Nanotechnol.* **2012**, *7*, 728-732.
4. Sun, P. Z.; Yang, Q.; Kuang, W. J.; Stebunov, Y. V.; Xiong, W. Q.; Yu, J.; Nair, R. R.; Katsnelson, M. I.; Yuan, S. J.; Grigorieva, I. V.; Lozada-Hidalgo, M.; Wang, F. C.; Geim, A. K. Limits on Gas Impermeability of Graphene. *Nature* **2020**, *579*, 229-232.
5. Todorovic, D.; Matkovic, A.; Milicevic, M.; Jovanovic, D.; Gajic, R.; Salom, I.; Spasenovic, M. Multilayer Graphene Condenser Microphone. *2D Mater.* **2015**, *2*, 045013.
6. Zhou, Q.; Zettl, A. Electrostatic Graphene Loudspeaker. *Appl. Phys. Lett.* **2013**, *102*, 223109.
7. Smith, A. D.; Niklaus, F.; Paussa, A.; Schroder, S.; Fischer, A. C.; Sterner, M.; Wagner, S.; Vaziri, S.; Forsberg, F.; Esseni, D.; Ostling, M.; Lemme, M. C. Piezoresistive Properties of Suspended Graphene Membranes under Uniaxial and Biaxial Strain in Nanoelectromechanical Pressure Sensors. *ACS Nano* **2016**, *10*, 9879-9886.
8. Lopez-Polin, G.; Ortega, M.; Vilhena, J. G.; Alda, I.; Gomez-Herrero, J.; Serena, P. A.; Gomez-Navarro, C.; Perez, R. Tailoring the Thermal Expansion of Graphene Via Controlled Defect Creation. *Carbon* **2017**, *116*, 670-677.
9. Abdullah, H. M.; Van der Donck, M.; Bahlouli, H.; Peeters, F. M.; Van Duppen, B. Graphene Quantum Blisters: A Tunable System to Confine Charge Carriers. *Appl. Phys. Lett.* **2018**, *112*, 213101
10. Lopez-Polin, G.; Jaafar, M.; Guinea, F.; Roldan, R.; Gomez-Navarro, C.; Gomez-Herrero, J. The Influence of Strain on the Elastic Constants of Graphene. *Carbon* **2017**, *124*, 42-48.
11. Suk, J. W.; Na, S. R.; Stromberg, R. J.; Stauffer, D.; Lee, J.; Ruoff, R. S.; Liechti, K. M. Probing the Adhesion Interactions of Graphene on Silicon Oxide by Nanoindentation. *Carbon* **2016**, *103*, 63-72.
12. Wood, J. D.; Harvey, C. M.; Wang, S. Adhesion Toughness of Multilayer Graphene Films. *Nat. Commun.* **2017**, *8*, 1952.
13. Lee, M.; Davidovikj, D.; Sajadi, B.; Siskins, M.; Alijani, F.; van der Zant, H. S. J.; Steeneken, P. G. Sealing Graphene Nanodrums. *Nano Lett.* **2019**, *19*, 5313-5318.
14. Ares, P.; Pisarra, M.; Segovia, P.; Diaz, C.; Martin, F.; Michel, E. G.; Zamora, F.; Gomez-Navarro, C.; Gomez-Herrero, J. Tunable Graphene Electronics with Local Ultrahigh Pressure. *Adv. Funct. Mater.* **2019**, *29*, 1806715.

15. Lee, C.; Wei, X. D.; Kysar, J. W.; Hone, J. Measurement of the Elastic Properties and Intrinsic Strength of Monolayer Graphene. *Science* **2008**, *321*, 385-388.
16. Lopez-Polin, G.; Gomez-Navarro, C.; Parente, V.; Guinea, F.; Katsnelson, M. I.; Perez-Murano, F.; Gomez-Herrero, J. Increasing the Elastic Modulus of Graphene by Controlled Defect Creation. *Nat. Phys.* **2015**, *11*, 26-31.
17. Boddeti, N. G.; Koenig, S. P.; Long, R.; Xiao, J.; Bunch, J. S.; Dunn, M. L. Mechanics of Adhered, Pressurized Graphene Blisters. *arXiv:1304.1011* **2020**.
18. Koenig, S. P.; G., B. N.; Dunn, L. M.; Bunch, J. S. Ultrastrong Adhesion of Graphene Membranes. *Nat. Nanotechnol.* **2011**, *6*, 543-546.
19. Choi, J. S.; Kim, J. S.; Byun, I. S.; Lee, D. H.; Lee, M. J.; Park, B. H.; Lee, C.; Yoon, D.; Cheong, H.; Lee, K. H.; Son, Y. W.; Park, J. Y.; Salmeron, M. Friction Anisotropy-Driven Domain Imaging on Exfoliated Monolayer Graphene. *Science* **2011**, *333*, 607-610.
20. Landau, L.; Lifshitz, E. *Fluid Mechanics*. Pergamon Press: 1987; Vol. 6.
21. Sutera, S. P.; Skalak, R. The History of Poiseuille's Law. *Annu. Rev. Fluid Mech.* **1993**, *25*, 1-20.
22. Vasic, B.; Matkovic, A.; Ralevic, U.; Belic, M.; Gajic, R. Nanoscale Wear of Graphene and Wear Protection by Graphene. *Carbon* **2017**, *120*, 137-144.
23. Bunch, J. S.; Dunn, M. L. Adhesion Mechanics of Graphene Membranes. *Solid State Commun.* **2012**, *152*, 1359-1364.
24. Gimeno, A.; Ares, P.; Horcas, I.; Gil, A.; Gomez-Rodriguez, J. M.; Colchero, J.; Gomez-Herrero, J. 'Flatten Plus': A Recent Implementation in Wsxm for Biological Research. *Bioinformatics* **2015**, *31*, 2918-2920.
25. Horcas, I.; Fernandez, R.; Gomez-Rodriguez, J. M.; Colchero, J.; Gomez-Herrero, J.; Baro, A. M. Wsxm: A Software for Scanning Probe Microscopy and a Tool for Nanotechnology. *Rev. Sci. Instrum.* **2007**, *78*, 013705.
26. Jaafar, M.; Martinez-Martin, D.; Cuenca, M.; Melcher, J.; Raman, A.; Gomez-Herrero, J. Drive-Amplitude-Modulation Atomic Force Microscopy: From Vacuum to Liquids. *Beilstein J. Nanotechnol.* **2012**, *3*, 336-344.
27. Sader, J. E. Frequency Response of Cantilever Beams Immersed in Viscous Fluids with Applications to the Atomic Force Microscope. *J. Appl. Phys.* **1998**, *84*, 64-76.
28. Sader, J. E.; Chon, J. W. M.; Mulvaney, P. Calibration of Rectangular Atomic Force Microscope Cantilevers. *Rev. Sci. Instrum.* **1999**, *70*, 3967-3969.
29. Markiewicz, P.; Goh, M. C. Atomic-Force Microscopy Probe Tip Visualization and Improvement of Images Using a Simple Deconvolution Procedure. *Langmuir* **1994**, *10*, 5-7.
30. Hertz, H. Über Den Kontakt Elastischer Körper. *J. Reine Angew. Mathematik* **1881**, *92*, 156-171.
31. Johnson, K. L. *Contact Mechanics*. Cambridge University Press: Cambridge, 1985.

# SUPPORTING INFORMATION

**SI1. Difference of the permeation rate for convex and concave configurations**

Leakage rate of pressurized graphene drumheads is usually lower in the convex (Figure S1a) than in the concave (Figure S1b) geometry. We attribute this difference to a higher effective adhesion of the membrane with the substrate when the pressure tends to push the membrane against the substrate. Additionally, the elastic energy of the deformed membrane is accumulated on the edges, causing a much higher interaction on this region when the pressure is higher outside the blister. In contrast, when the pressure is higher inside the blister, it will tend to detach the membrane, even causing delamination/peeling of the flake when the pressure is high enough.[1, 2] The difference on the characteristic times for the two geometries varies significantly between different drumheads, but it is around a factor of 1.5 to 3.

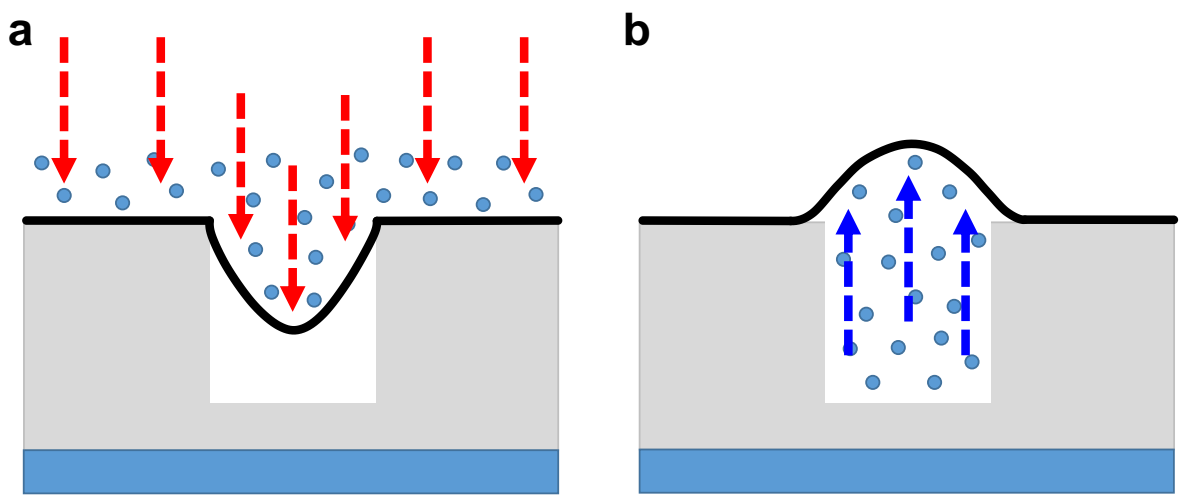

**Figure S1.** Origin of the difference between the permeation rate for convex (a) and concave (b) blister configurations. For the concave case, if the pressure inside the blister is high enough, it can even produce delamination/peeling of the membrane, facilitating the exit of the gas molecules through the membrane-substrate interface. On the contrary, this will not occur in the convex case, where the higher pressure outside the blister will increase effective adhesion of the membrane.

## SI2. Gas diffusion through the graphene-SiO$_2$ interface

A simple but useful analogy for the behaviour of the gas leakage through the graphene-SiO$_2$ interface is the gas flow through a cylindrical pipe. As represented in Figure S2, the flow is dependent on the length and the cross section of the tube. Sealing the cavities with a diamond tip is comparable to constrict the tube in some parts, consequently diminishing the leak.

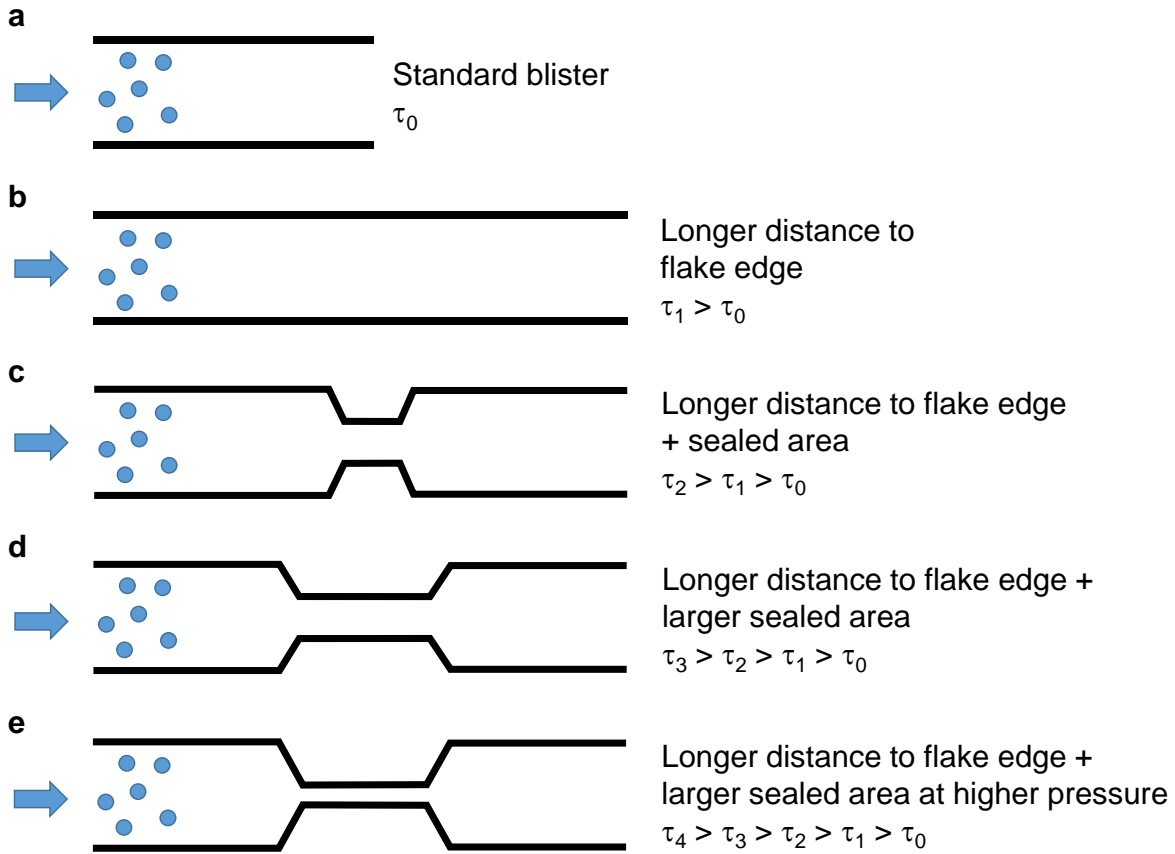

**Figure S2.** Analogy of the diffusion of gas molecules from the graphene blisters with the diffusion through channels of different length and width. (a) Standard blister. (b) Standard blister with a longer distance to the flake edge. (c) Similar as (b) but with a sealed area around the blister. (d) Similar as (c) but with a larger sealed area. (e) Similar as (d) but with the area sealed using higher pressure.

Poiseuille's law gives the pressure drop in a fluid flowing through a long cylindrical pipe of constant cross section.[3, 4] For a compressible gas, if one of the sides of the tube of radius $r$ and length $L$ is in high vacuum ($P\sim 0$), the equation describing the volumetric flow rate ($Q$) through the pipe is

$$Q = \frac{dV}{dt} = \frac{1}{2}\frac{\pi r^4}{8\eta L}\Delta P \quad (1)$$

$$\frac{dV}{dt} = \frac{1}{\rho}\frac{dm}{dt} = \frac{M}{\rho}\frac{dn}{dt} = \frac{MV}{\rho RT}\frac{dP}{dt} \quad (2)$$

Where $V$ is the volume, $P$ the pressure, $\rho$ the gas density (1.18 kg m$^{-3}$ for air at 298 K), $m$ is the mass, $M$ is the molar mass of the gas (~ 0.029 kg mol$^{-1}$ for air) and $\eta$ is the dynamic viscosity of the fluid (1.85×10$^{-5}$ kg m$^{-1}$ s$^{-1}$ for air at 298 K). Combining equations (1) and (2), we get

$$\frac{dP}{dt} = -\frac{\rho RT \pi r^4}{16 \eta MLV} P \quad (3)$$

Then

$$P = Ce^{-kt} \quad (4)$$

$$k = 1/\tau = \frac{\rho RT \pi r^4}{16 \eta MLV} \quad (5)$$

With $\tau$ the characteristic leakage time. For the geometry of a circular membrane, the pressure difference $\Delta P$ as a function of the deflection $Z$ can be expressed as[5]

$$\Delta P = \frac{4}{\pi a^2}\left(c_1 S_0 Z + \frac{4 c_2 Et}{\pi a^2 (1-\nu)} Z^3\right) \quad (6)$$

Where $a$ is the radius of the pressurized circular region, $c_1$ = 3.393, $c_2$ = (0.8 + 0.062 $\nu$)$^{-3}$ with $\nu$ the Poisson's ratio (~ 0.16), $Et$ = 340 N m$^{-1}$ for monolayer graphene and $S_0$ is the pre-tension accumulated in the sheet. This pre-tension can be very variable from membrane to membrane, with typical values ranging from 0.05 to 0.8 N m$^{-1}$.[6] Thus, we consider two types of scenarios: membranes dominated by the pre-tension, where $\Delta P \sim Z$ (valid for small deflections), and membranes dominated by stretching, where $\Delta P \sim Z^3$.

In our measurements, the maximum deflection values are relatively small (typically below 20 nm), so we can initially consider $\Delta P \sim Z$ and hence

$$Z \approx \frac{\pi a^2}{4 c_1 S_0} P \quad (7)$$

Then

$$Z \approx \frac{\pi a^2}{4 c_1 S_0} C e^{-kt} \Rightarrow \boldsymbol{Z \approx A e^{-kt}} \quad (8)$$

With

$$A = \frac{\pi a^2}{4 c_1 S_0} C \quad (9)$$

$$\tau = 1/k = \frac{16 \eta MV}{\pi \rho RT} \frac{L_{eff}}{r_{eff}^4} \quad (10)$$

As can be seen, the dependence of the height of the blisters follows the same exponential decay we heuristically fitted in the manuscript, with $L_{eff}$ the distance of the blister to the flake edge and $r_{eff}$ the effective radius of a tube that yields the observed leakage rate.

If we now consider the scenario $\Delta P \sim Z^3$, then

$$P \approx \frac{4}{\pi a^2} \frac{4c_2 Et}{\pi a^2(1-v)} Z^3 = Ce^{-kt} \Rightarrow Z \approx \left(\frac{\pi^2 a^4(1-v)}{16 c_2 Et} C\right)^{1/3} e^{-\frac{k}{3}t} \Rightarrow \boldsymbol{Z \approx A' e^{-k't}}$$

Where $\quad A' = \left(\frac{\pi^2 a^4(1-v)}{16 c_2 Et} C\right)^{1/3}$ and $k' = k/3$

As can be seen, the dependence of the height of the blisters again follows a similar exponential decay.

This toy model yields valuable and still intuitive information of the system. At first glance, this analysis accounts for the linear increase of the blisters gas leakage times with the distance to the flake edge, as observed experimentally (Figure 2d in the main text). The analysis also allows us to obtain values for the effective radius of the leakage, and thus to better compare leakage rates from blisters in different flakes and leakages before and after sealing. From this analogy, we obtain effective radii for the gas leakage ranging from 0.3 to 1.7 nm before sealing the blisters. Please note that in the scenario where $\Delta P \sim Z^3$, the characteristic time would be $\tau' = 1/k' = 3\tau$. As $r_{eff} \sim \tau^{1/4}$, the differences in the effective radius values obtained from the two scenarios will not be very significant.

**SI3. The role of graphene-substrate interaction on the characteristic times of blisters**

Together with the distance of the blister to the flake edge or the size of the blister, other effects such as the uncontrolled initial interaction after transfer between graphene and substrate have a strong influence on the inflation/deflation time of the blisters. We performed the experiments described in the manuscript on tens of membranes, getting extremely different results. Figure S3a shows the characteristic times obtained for blisters of similar characteristics (radii *a* between 500 and 750 nm and distances to the flake edge between 2.0 and 2.5 μm) in five different significant samples. It is remarkable that the blisters from samples 4 and 5, where the distances to the flake edge were much higher, 4 and 7 μm respectively, and thus higher times would be expected, exhibit the lowest characteristic times of the plot. Consequently, we attribute this disparity to uncontrolled variations of the graphene-$SiO_2$ adhesion between different flakes. This is as well reflected on the different effective leakage radii obtained for each of the different samples (Figure S3b).

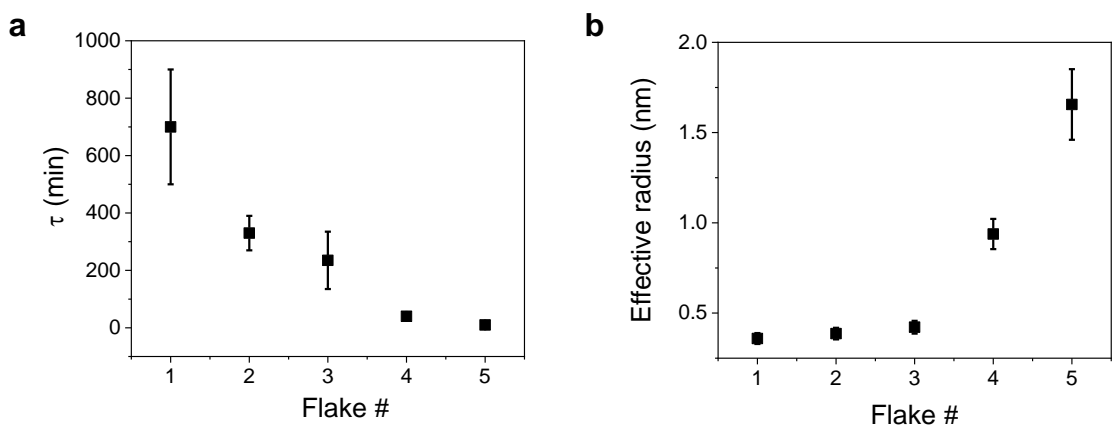

**Figure S3.** (a) Characteristic times and (b) effective leakage radii of blisters of 5 different samples. The disparity is not related to the distance to the edge, as described above.

## SI4. Dimensions of sealed areas

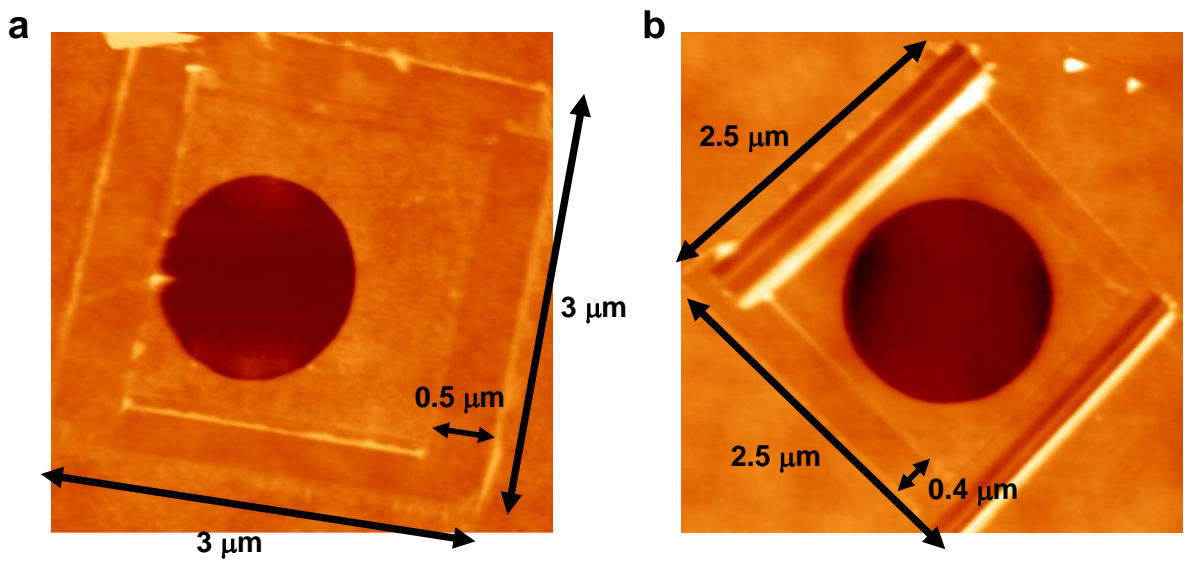

**Figure S4.** Topographic images of two different sealed blisters with the characteristic dimensions of the sealed areas.

## SI5. Increase of graphene-substrate interaction after sealing

The reduction of leakage through the graphene-substrate interface after sealing indicates an increase of the graphene-substrate interaction. This increase upon sealing of drumheads can be very useful for a variety of situations. The total force that a membrane can withstand before breaking under indentation with a tip depends strongly on the radius of the tip. Consequently, the maximum strain produced on the membrane during an indentation is also strongly influenced by the tip radius. Additionally, graphene-substrate interaction can be an issue to achieve high strains, as the membranes tend to sag when applying very high forces.

To gain a qualitative insight on the increase of the interaction between the graphene membrane and the underlying $SiO_2$ substrate upon sealing, we performed indentations on sealed and unsealed graphene drumheads using tips with a radius of 250 nm. These especial tips allow us to achieve high average strains necessary to overcome the flake-substrate interaction without breaking the membrane Figure S5a and Figure S5b show images of both drumheads before indentation, where no significant defects such as wrinkles are visible. When indenting the unsealed drumhead at low indentation (Figure S5c), both approach and retract curves overlap, indicating an elastic response of the membrane. For high enough indentations, we can observe a clear hysteresis between the approach and the retract curves (Figure S5e), which is not observed in the sealed case for the same (and even much higher) indentation (Figure S5d). This hysteresis is related to the slippage of the membrane edges during the indentation process. This observation evidences that the sealed drumhead exhibits a much higher interaction with the substrate than the unsealed one. By performing deeper indentations (Figure S5f), we also observe a hysteretic behaviour even in the sealed membrane. AFM images acquired after hysteretic indentations show marked wrinkles in the supported graphene (Figure S5g and Figure

S5h), which were not there before, and thus they are the fingerprint of the slippage of the membrane. Our data show that the hysteresis onset occurs at forces almost 3 times higher for the sealed case, confirming a much higher membrane-substrate interaction upon ultrahigh pressure sealing.

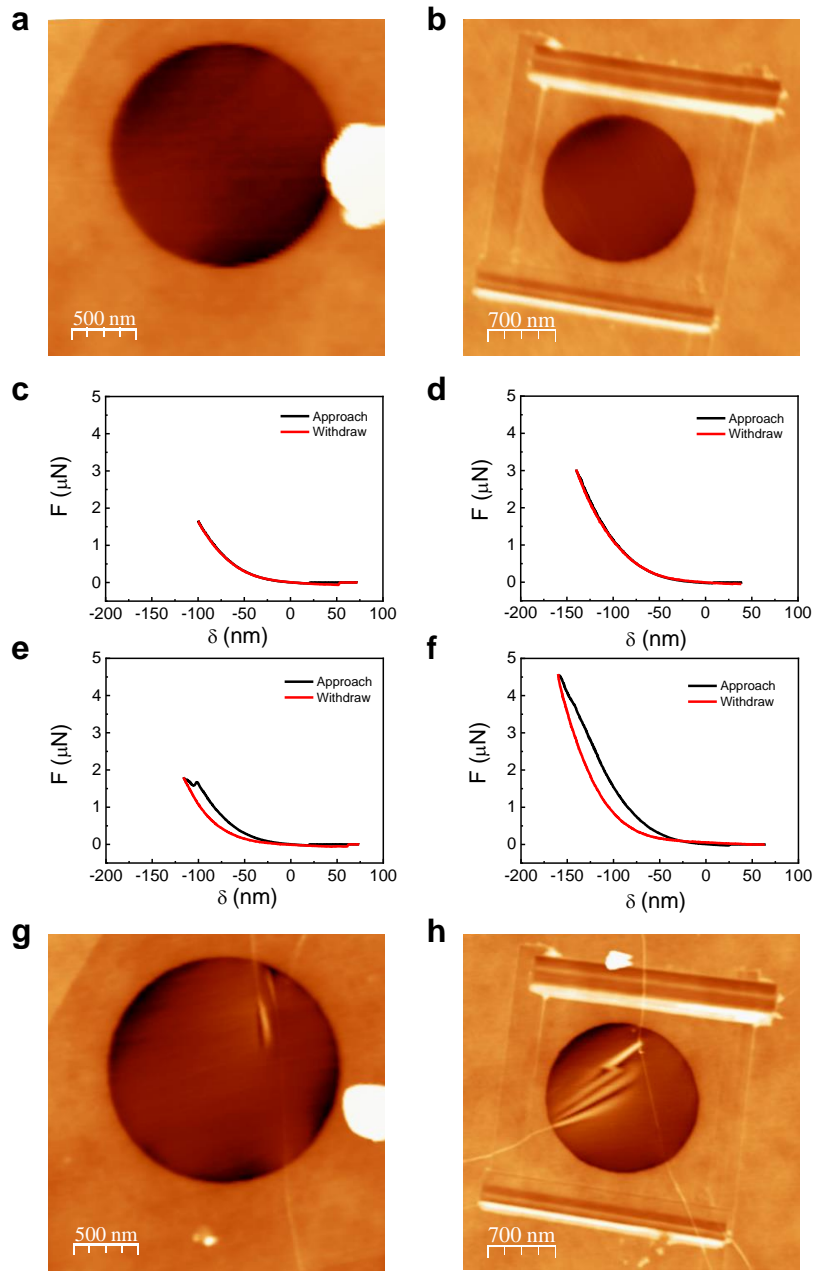

**Figure S5.** Unsealed (a) and sealed (b) drumheads before high force indentations. Indentation curves for the unsealed (c,e) and sealed (d,f) drumheads. (g) Unsealed and (h) sealed drumheads after high force indentations. Indentation curves are plotted with the same scales for a better direct comparison.

**References**

1. Boddeti, N. G.; Koenig, S. P.; Long, R.; Xiao, J.; Bunch, J. S.; Dunn, M. L. Mechanics of Adhered, Pressurized Graphene Blisters. *arXiv:1304.1011* **2020**.
2. Koenig, S. P.; G., B. N.; Dunn, L. M.; Bunch, J. S. Ultrastrong Adhesion of Graphene Membranes. *Nat. Nanotechnol.* **2011**, *6*, 543-546.
3. Landau, L.; Lifshitz, E. *Fluid Mechanics*. Pergamon Press: 1987; Vol. 6.
4. Sutera, S. P.; Skalak, R. The History of Poiseuille's Law. *Annu. Rev. Fluid Mech.* **1993**, *25*, 1-20.
5. Bunch, J. S.; Verbridge, S. S.; Alden, J. S.; van der Zande, A. M.; Parpia, J. M.; Craighead, H. G.; McEuen, P. L. Impermeable Atomic Membranes from Graphene Sheets. *Nano Lett.* **2008**, *8*, 2458-2462.
6. Lopez-Polin, G.; Gomez-Navarro, C.; Parente, V.; Guinea, F.; Katsnelson, M. I.; Perez-Murano, F.; Gomez-Herrero, J. Increasing the Elastic Modulus of Graphene by Controlled Defect Creation. *Nat. Phys.* **2015**, *11*, 26-31.